\documentclass[10pt,aps,prl,amsmath,amssymb,superscriptaddress,floatfix,nofootinbib,notitlepage,reprint]{revtex4-1}
\usepackage{amsmath,amssymb}
\usepackage{graphicx,color}
\usepackage[colorlinks=true,hyperfootnotes=false]{hyperref}
\usepackage{epstopdf}
\usepackage{txfonts}

\begin{document}
\title{\boldmath Can $X(5568)$ be described as a $B_s\pi$, $B\bar{K}$ resonant state?}
\author{Miguel Albaladejo}
\affiliation{Instituto de F\'isica Corpuscular (IFIC),
             Centro Mixto CSIC-Universidad de Valencia,
             Institutos de Investigaci\'on de Paterna,
             Aptd. 22085, E-46071 Valencia, Spain}
\author{Juan Nieves}
\affiliation{Instituto de F\'isica Corpuscular (IFIC),
             Centro Mixto CSIC-Universidad de Valencia,
             Institutos de Investigaci\'on de Paterna,
             Aptd. 22085, E-46071 Valencia, Spain}
\author{Eulogio Oset}
\affiliation{Instituto de F\'isica Corpuscular (IFIC),
             Centro Mixto CSIC-Universidad de Valencia,
             Institutos de Investigaci\'on de Paterna,
             Aptd. 22085, E-46071 Valencia, Spain}
\author{Zhi-Feng Sun}
\affiliation{Instituto de F\'isica Corpuscular (IFIC),
             Centro Mixto CSIC-Universidad de Valencia,
             Institutos de Investigaci\'on de Paterna,
             Aptd. 22085, E-46071 Valencia, Spain}
\author{Xiang Liu}
\affiliation{School of Physical Science and Technology, Lanzhou University, Lanzhou 730000, China}
\affiliation{Research Center for Hadron and CSR Physics, Lanzhou University and Institute of Modern Physics of CAS, Lanzhou 730000, China}

\begin{abstract}
The D0 Collaboration has recently seen a resonant-like peak in the $B_s\pi$ invariant mass spectrum, claimed to be a new state called $X(5568)$. Using a $B_s\pi$--$B\bar{K}$ coupled channel analysis, implementing unitarity, and with the interaction derived from Heavy Meson Chiral Perturbation Theory, we are able to reproduce the reported spectrum, with a pole that can be associated to the claimed $X(5568)$ state, and with mass and width in agreement with the ones reported in the experimental analysis. However, if the $T$-matrix regularization is performed by means of a momentum cutoff, the value for the latter needed to reproduce the spectrum is $\Lambda = 2.80 \pm 0.04\ \text{GeV}$, much larger than a ``natural'' value $\Lambda \simeq 1\ \text{GeV}$. In view of this, it is difficult to interpret the nature of this new state. This state would not qualify as a resonance dynamically generated by the unitarity loops. Assuming the observed peak to correspond to a physical state, we make predictions for partners in the $D$, $D^\ast$, and $B^\ast$ sectors. Their observation (or lack thereof) would shed light into this issue.
\end{abstract}

\maketitle

\paragraph{Introduction.---} The D0 Collaboration has recently claimed \cite{D0:2016mwd} the discovery of a new state, $X(5568)$, seen as a clear peak in the $B^0_s \pi^{\pm}$ invariant mass spectrum, with a mass $M_X = 5567.8 \pm 2.9^{+0.9}_{-1.9}\ \text{MeV}$ and a width $\Gamma_X = 21.9 \pm 6.4^{+5.0}_{-2.5}\ \text{MeV}$. The spectroscopy of mesons and baryons with heavy constituent quarks is living an exciting era, in which every so often new states are announced \cite{Chen:2016qju,Olsen:2014qna}. If confirmed, this new $X(5568)$ state would be even more exotic, for it would be the first one involving four different flavour quarks ---its quark content would be $\bar{b}s\bar{d}u$ (for the $B^0_s \pi^+$ case). Given the ``exoticness'' of this state, the announcement has been followed by a large number of theoretical papers \cite{Agaev:2016mjb,Wang:2016tsi,Wang:2016mee,Zanetti:2016wjn,Chen:2016mqt,Liu:2016xly,Agaev:2016ijz,Liu:2016ogz,Agaev:2016lkl,Dias:2016dme, Wang:2016wkj,Agaev:2016urs,He:2016yhd,Jin:2016cpv,Stancu:2016sfd,Burns:2016gvy,Tang:2016pcf,Guo:2016nhb,Lu:2016zhe,Esposito:2016itg} trying to explain its properties or its nature, or to point out the difficulties one encounters in its description. Even more recently, the LHCb Collaboration has presented preliminary results \cite{LHCb:2016ppf} for the spectrum for the same final state, with negative results.

In this work, we present a $B^0_s \pi$, $B\bar{K}$ coupled channel $T$-matrix analysis, with the aim of reproducing for the first time the $B^0_s \pi$ invariant mass spectrum reported by the D0 Collaboration \cite{D0:2016mwd}, in which the $X(5568)$ peak has been seen. A fit to the spectrum is done, and the only free parameter of the $T$-matrix fixes the energy position of the peak, and its width comes out naturally. Yet, we shall make a discussion on how to interpret this result to the light of the value obtained for this parameter used to render finite the unitarized $T$-matrix.

\begin{figure}[!t]
\includegraphics[height=3.3cm,keepaspectratio]{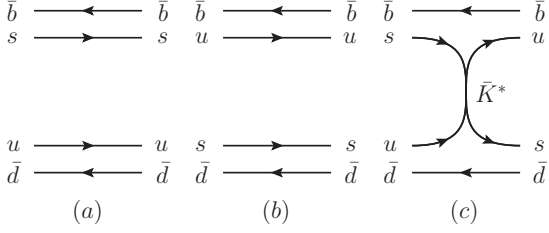}
\caption{Diagramatic representation in terms of quarks of the interactions $B^0_s\pi^+ \to B^0_s\pi^+$ (a), $B^+ \bar{K}^0 \to B^+ \bar{K}^0$ (b) and $B^0_s \pi^+ \to B^+ \bar{K}^0$ (c).\label{fig:quarks}}
\end{figure}

\paragraph{Coupling the $B^0_s\pi$, $B \bar{K}$ states.---} The $X(5568)$ has been found as a peak in the invariant mass spectrum of the states $B_s\pi^{\pm}$, which are $I=1$, $I_z = \pm 1$ states, which should couple, in principle, to the $I=1$ state $B^+ \bar{K}^0$, for $I_z=+1$, or $B^0 K^-$, for $I_z=-1$. We consider the $I(J^P) = 1(0^+)$ coupled channels $B_s\pi$ (1) and $B\bar{K}$ (2), for which the $T$-matrix can be written as:
\begin{equation}
T^{-1}(s) = V^{-1}(s) - G(s)~,
\end{equation}
where the diagonal matrix $G$ contains the two-meson one-loop functions for $B_s\pi$ and $B\bar{K}$, and $V$ is a symmetric matrix with matrix elements given by the transition potentials between the two channels, to be discussed later. The Mandelstam variable $s=M_{B_s\pi}^2$ is the $B_s\pi$ center-of-mass energy squared. The loop functions are regularized in this work by means of a subtraction constant,
\begin{align}
& 16\pi^2 G_i(s) = a_i(\mu) +  \log\frac{M_i m_i}{\mu^2} + \frac{\Delta_i}{2s}\log\frac{M_i^2}{m_i^2} \nonumber  \\
& + \frac{\nu_i}{2s} 
\left( 
\log\frac{s-\Delta_i+\nu_i}{-s+\Delta_i+\nu_i} + 
\log\frac{s+\Delta_i+\nu_i}{-s-\Delta_i+\nu_i}
\right) \label{eq:GloopSubtracted}~, \\
& \Delta_i = M_i^2-m_i^2~, \quad \nu_i = \lambda^{1/2}(s,M_i^2,m_i^2)~, \nonumber
\end{align}
where $M_i$ ($m_i$) is the mass of the heavy (light) meson in the $i$th channel.  We fix $\mu = (M_1 + M_2)/2$, and, although the two subtraction constants $a_i(\mu)$ could be in principle different, we set $a_1(\mu)=a_2(\mu)=a(\mu)$. We have checked that no significant differences are found if this constraint is not imposed. The $V$ matrix elements are computed by means of Heavy Meson Chiral Perturbation Theory \cite{Wise:1992hn,Manohar:2000dt}. Interestingly, one finds that the only non-zero $S$-wave potential is the off-diagonal one:
\begin{align}
V_{11}(s) & = V_{22}(s) = 0~,\label{eq:pot-diag} \\
V_{12}(s) & = \frac{1}{8f^2} \left( 3s -\left(M_1^2+M_2^2+m_1^2+m_2^2\right) - \frac{\Delta_1 \Delta_2}{s} \right)~, \label{eq:pot-offdiag}
\end{align}
where $f \simeq 93\ \text{MeV}$ is the pion weak decay constant. The same results for these matrix elements are obtained if one uses instead the Local Hidden Gauge approach \cite{Bando:1987br,Meissner:1987ge}, where the interaction is driven by the exchange of vector mesons. We see in Fig.~\ref{fig:quarks} that the diagonal matrix elements are zero, as in Eq.~\eqref{eq:pot-diag}, since no $q\bar{q}'$ pair can be exchanged between $B^0_s$ and $\pi^+$ [Fig.~\ref{fig:quarks}(a)], nor between $B^+$ and $\bar{K}^0$ [Fig.~\ref{fig:quarks}(b)]. The interaction is OZI forbidden. However, as seen in Fig.~\ref{fig:quarks}(c), the $B^0_s \pi^+ \to B^+ \bar{K}^0$ transition is allowed through the exchange of an $s\bar{u}$ pair forming a $K^{\ast-}$. Computing this diagram in the aforementioned local hidden gauge approach renders the same potential as in Eq.~\eqref{eq:pot-offdiag}.

The structure of the kernel matrix $V$ will be very important in the interpretation of $X(5568)$. For the $T$-matrix, it implies the structure:
\begin{equation}
T(s) = \frac{V_{12}(s)}{D(s)} \left(\begin{array}{cc} G_2(s) V_{12}(s) & 1 \\ 1 & G_1(s) V_{12}(s) \end{array} \right)~,
\end{equation}
where $D(s) = 1 - V_{12}(s)^2 G_1(s) G_2(s)$. If our formalism is able to reproduce the $X(5568)$ resonant peak, a pole should appear on the second Riemann sheet of the $T$-matrix at the position $s = s_X$, where $\sqrt{s_X} \equiv M_X - i \Gamma_X/2$,
\begin{equation}
T_{ij}(s) \simeq \frac{\xi_i \xi_j}{s-s_X} + \cdots~,
\end{equation}
where $\xi_i$ is the coupling of $X(5568)$ to the $i$th channel. This implies a zero in the second Riemann sheet of the $D(s)$ function, $D^{(II)}(s_X) = 0$.


\begin{figure}[t!]\centering
\includegraphics[height=1.7cm,keepaspectratio]{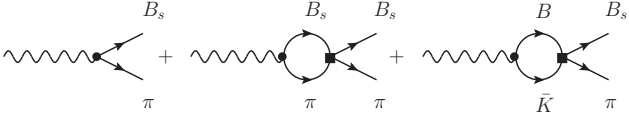}
\caption{$B^0_s \pi$ Production mechanism. The wavy line represents a generic source with $B^0_s\pi$ quantum numbers.\label{fig:diag}}
\end{figure}

\paragraph{Invariant $B^0_s\pi$ mass spectrum.---} For the $B_s\pi$ production mechanism, we write a generic amplitude (see Fig.~\ref{fig:diag}),
\begin{equation}
t(s) = f_1\ \bigg( 1 + G_1(s) T_{11}(s) \bigg) + f_2\ G_2 T_{21}(s)~,
\end{equation}
in which the two couplings $f_i$ are unknown. Actually, because in the $B_s \pi$ spectrum there is an unknown global normalization constant, the only relevant quantity is the ratio $f_1/f_2$. We can consider these couplings to be proportional to the couplings of $X(5568)$ to $B_s\pi$ and $B\bar{K}$, and hence:
\begin{equation}
\frac{f_1}{f_2} \simeq \frac{\xi_1}{\xi_2} = V_{12}(s_X) G_2(s_X)~.
\end{equation}
In this way, we can write down the amplitude $t(s)$ without any new free parameter, except for an irrelevant global constant, as:
\begin{equation}
t(s) = f_2 \frac{V_{12}(s) G_2 (s)}{D(s)} \left( 1 + \frac{V_{12}(s_X)G_2(s_X)}{V_{12}(s) G_2(s)} \right)~.
\end{equation}
Notice that what really matters in this production amplitude is that it is proportional to the function $1/D(s)$, where the $X(5568)$ pole will show up.

Finally, the $B_s \pi$ invariant spectrum has the form:
\begin{align}
N(s) & = \frac{p(s)}{\sqrt{s}} \left( N_\text{amp}\ |t(s)|^2 + N_\text{bkg}\ f_\text{bkg}(s) \right)~,\label{eq:spec}\\
p(s) & = \frac{\lambda^{1/2}\left(s,M_{B_s}^2,m_\pi^2 \right)}{2\sqrt{s}}~, \nonumber
\end{align}
where the background $f_\text{bkg}(s)$ is parameterized as: 
\begin{equation}\label{eq:bkg}
f_\text{bkg}(s) = P_A(s)\ \text{exp}\big(P_B(s)\big)~,
\end{equation}
similarly as done in the experimental analysis \cite{D0:2016mwd}. The functions $P_{A,B}(s)$ are polynomials in the variable $x_s = (s-s_\text{th})/s_\text{th}$, $s_\text{th} = (M_1 + m_1)^2$, with free coefficients $C_{A,i}$, $C_{B,i}$:
\begin{align}
P_A(s) & = 1 + C_{A,2}\ x_s^2 + C_{A,3}\ x_s^3~,\label{eq:polyA}\\
P_B(s) & = C_{B,1}\ x_s + C_{B,2}\ x_s^2 C_{B,3}\ x_s^3~.\label{eq:polyB}
\end{align}

\begin{figure}[t!]
\includegraphics[height=5.2cm,keepaspectratio]{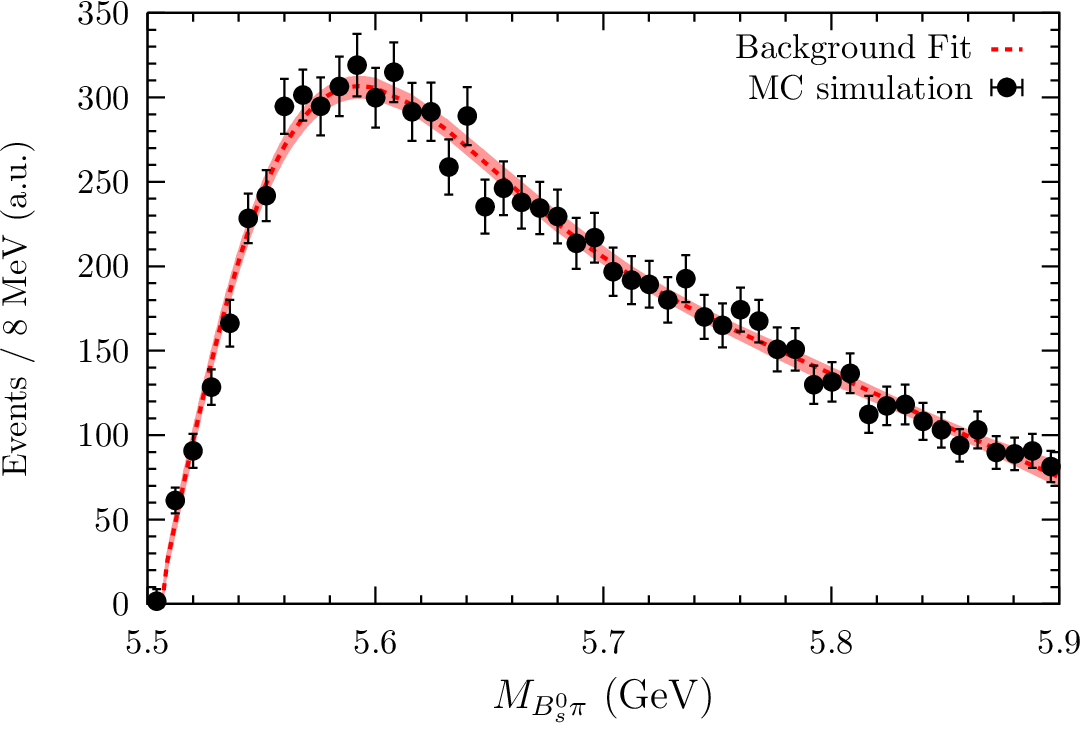}\\
\includegraphics[height=5.2cm,keepaspectratio]{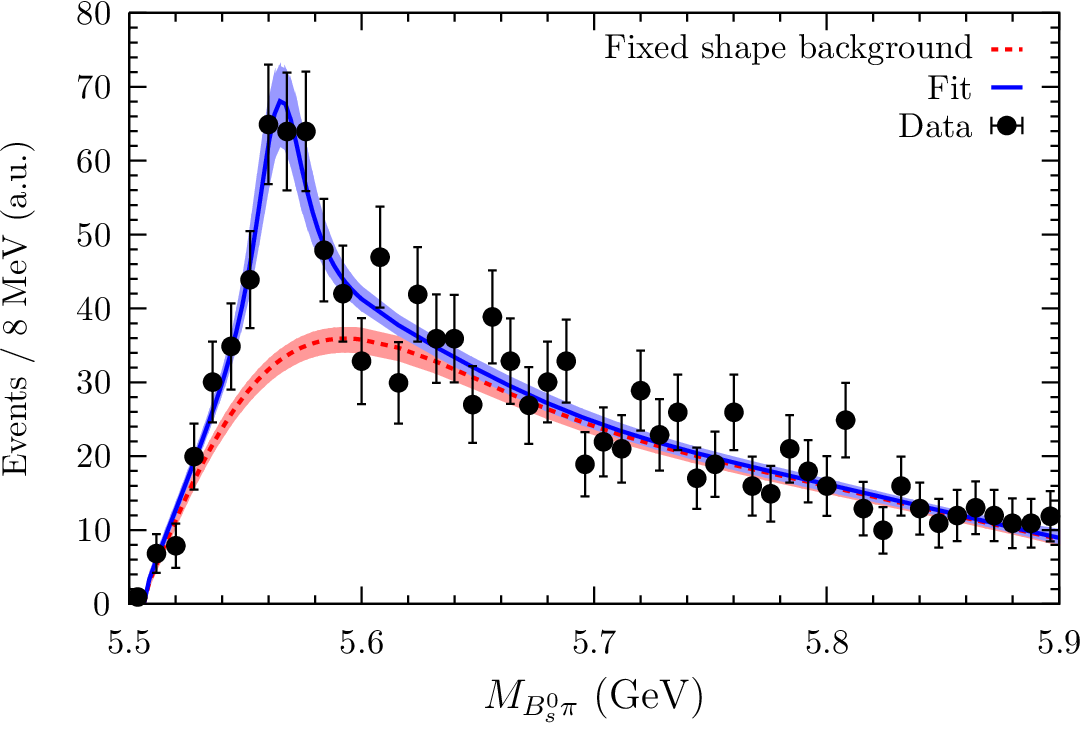}
\caption{Top: The points represent the MC simulation reported by the D0 Collaboration \cite{D0:2016mwd} for the background of the $B^0_s\pi$ invariant mass spectrum. The dashed red line represents our fit, the shape of which is given by Eq.~\eqref{eq:bkg}. Bottom: The points represent the $B^0_s \pi$ invariant mass spectrum measured by the D0 Collaboration \cite{D0:2016mwd}. The solid blue line represent our fit of Eq.~\eqref{eq:spec} to these data. The background shape is fixed to that of the top panel. The MC simulation (top) and the spectrum (bottom) refer to the ones reported by the D0 Collaboration when the ``cone cut'' criterion is applied. See text and Ref.~\cite{D0:2016mwd} for details. In both panels, our statistical errors, given by the error bands, represent $1\sigma$ confidence intervals, and are estimated by Monte Carlo resampling of the data \cite{Press:1992zz}.\label{fig:spec}}
\end{figure}
\begin{figure}[t!]
\includegraphics[height=5.2cm,keepaspectratio]{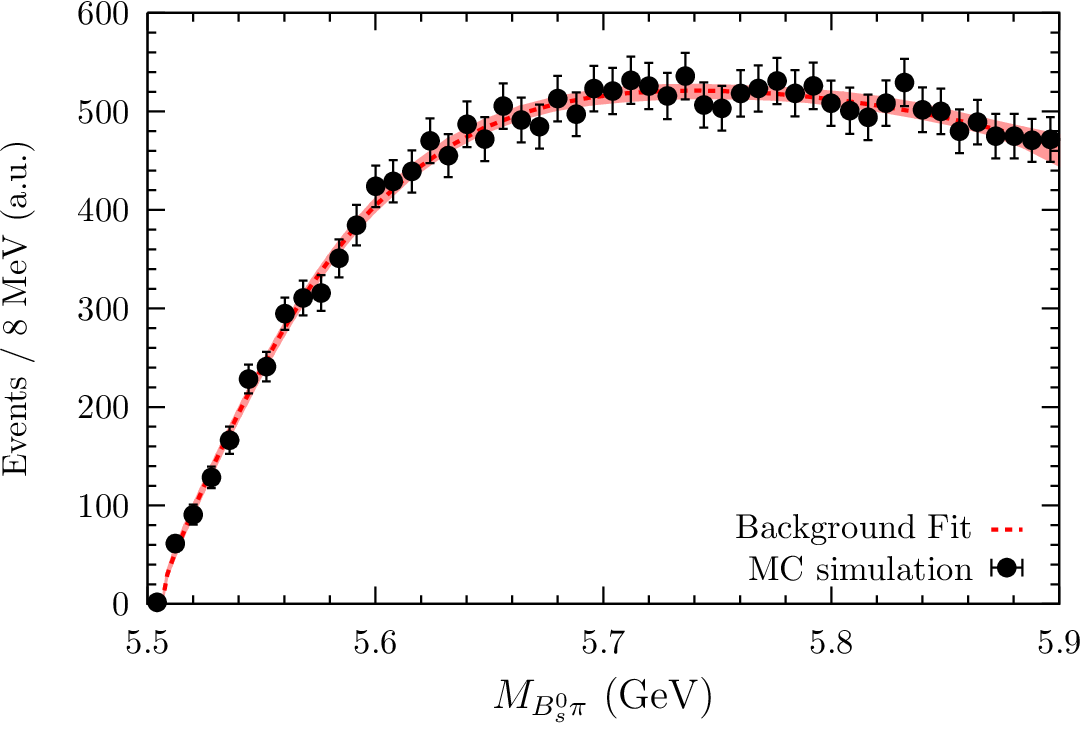}\\
\includegraphics[height=5.2cm,keepaspectratio]{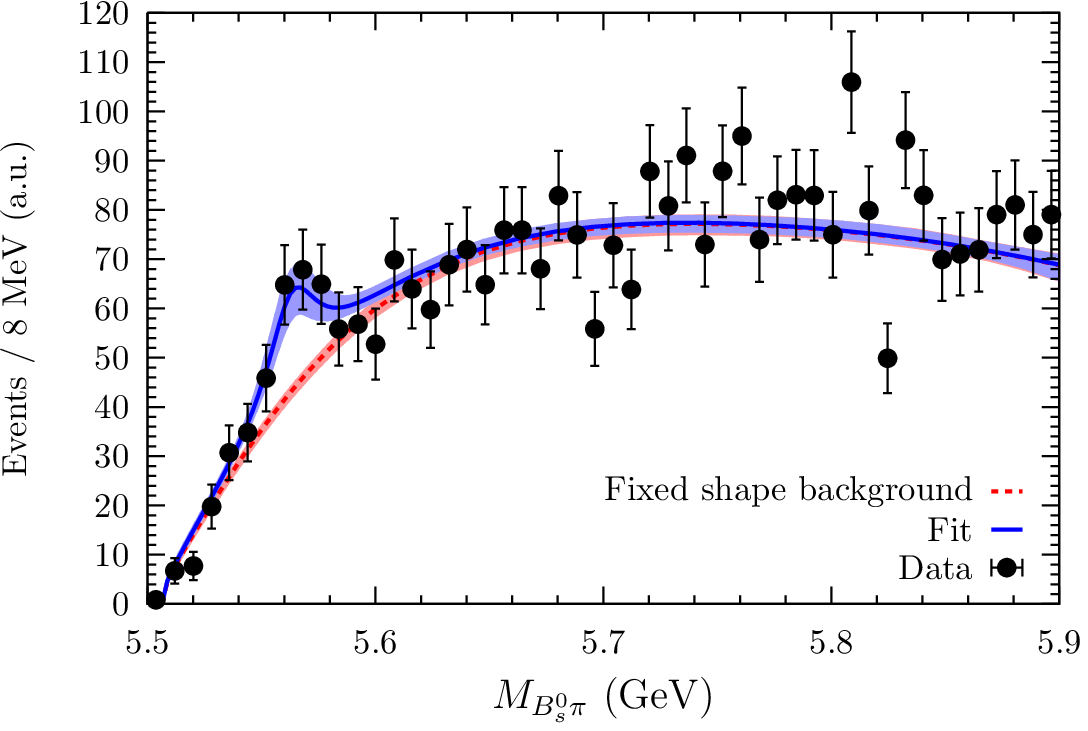}
\caption{Same as Fig.~\ref{fig:spec}, but for the case in which the ``cone-cut'' criterion is not applied. See text and Ref.~\cite{D0:2016mwd} for details.\label{fig:spec_noconecut}}
\end{figure}

\paragraph{Results and direct interpretation.---} The D0 Collaboration has reported the $B^0_s \pi^\pm$ spectrum coming from $p\!\bar{p}$ collisions data at $1.96\ \text{TeV}$. They report two different cases, depending on wether the ``cone cut'' criterion is applied or not in the events selection (see Ref.~\cite{D0:2016mwd} for further details). Without entering into the details, this criterion, besides reducing the background, it clearly enhances the $X(5568)$ peak region. For completeness, we shall study both cases.
 
Similarly as done in Ref.~\cite{D0:2016mwd}, for each of the spectra (with and without the ``cone cut'') we perform a two-step fit to the $B_s\pi$ invariant mass spectrum. In the first step, the shape of the background is fixed by fitting the function $f_\text{bkg}$ [Eq.~\eqref{eq:bkg}] to the MC simulation {\it data} given in Ref.~\cite{D0:2016mwd}. This step fixes then the coefficients of the polynomials in Eqs.~\eqref{eq:polyA} and \eqref{eq:polyB}. The result of this first step is shown in the top panels of Figs.~\ref{fig:spec} and \ref{fig:spec_noconecut} for the cases with and without ``cone cut'', respectively. Good agreement is seen between the MC simulation and our fit. On a second step, with the background shape already fixed, we fit the theoretical spectrum of Eq.~\eqref{eq:spec} to the experimental invariant mass distribution \cite{D0:2016mwd}. This second step requires fitting the two normalization constants $N_\text{amp}$ and $N_\text{bkg}$, together with the only free parameter entering in our $T$ matrix, namely, the subtraction constant $a(\mu)$, for which we find $a(\mu) = -0.97 \pm 0.02$, for the ``cone cut'' spectrum, and $a(\mu)=-0.98 \pm 0.02$ for the spectrum without the ``cone cut''. These fits are shown in the bottom panels of Figs.~\ref{fig:spec} and \ref{fig:spec_noconecut}, respectively, with a blue solid line. The agreement is excellent, and, indeed, the fits have a $\chi^2_\text{d.o.f.} = 38/(50-3) = 0.8$ and $66/(50-3) = 1.4$, respectively. The latter is larger, mainly due to the large variations of the tail of the spectrum. From this fit, the $X(5568)$ parameters are found to be $M_X = 5564.2 \pm 2.6\ \text{MeV}$ and $\Gamma_X = 26.7 \pm 1.2\ \ \text{MeV}$ for the ``cone cut'' spectrum, and $M_X = 5562.8 \pm 2.8\ \text{MeV}$ and $\Gamma_X = 27.4 \pm 1.2\ \text{MeV}$, in good agreement with the experimental determination of Ref.~\cite{D0:2016mwd}. 

In our formalism, the $X(5568)$ appears as a pole in the $B^0_s \pi$, $B \bar{K}$ coupled channel unitary $T$-matrix. However, since the diagonal terms of the potential matrix $V$ are zero, the whole interaction is driven by the off-diagonal potential $V_{12}(s)$, that connects both channels. This means that both channels are strictly necessary to originate the pole. This can be particularly well seen by noticing that the pole condition, $D^{(II)}(s_X) = 0$, cannot be achieved if any of $V_{12}$, $G_1$ or $G_2$ are set to zero, because $D(s) = 1 - V_{12}(s)^2 G_1(s) G_2(s)$. In this sense, one cannot say that the $X(5568)$ is a purely $B\bar{K}$ bound state nor a purely $B_s\pi$ resonant state, but a resonant state made out of both channels.

\paragraph{Further discussions.---} As pointed out in the introduction, the LHCb Collaboration \cite{LHCb:2016ppf} has reported preliminary negative results in the search for this state in the $B_s\pi$ spectrum produced in $p\!p$ collisions at $7$ and $8\ \text{TeV}$, whereas the spectrum reported by the D0 Collaboration \cite{D0:2016mwd} is originated in $p\!\bar{p}$ collisions at $1.96\ \text{TeV}$. Although the mechanisms involved in the $B_s\pi$ production are different in LHCb and D0, this should not be important given the large $p\!p$ or $p\!\bar{p}$ energies involved in both cases. Hence, this disagreement is, in principle, unexpected. The LHCb work explicitly states that the ``cone cut'' selection criterion can generate broad peaking structures. Indeed, a broad peak in the background at $\sqrt{s} \simeq 5.6\ \text{GeV}$ can be seen in the top panel of Fig.~\ref{fig:spec}. However, we have also seen that the $X(5568)$ structure is present regardless of wether this criterion is imposed (Fig.~\ref{fig:spec}) or not (Fig.~\ref{fig:spec_noconecut}). On the other hand, the presence of this structure is much less clear if the ``cone cut'' is not applied.

One should remark, as a non trivial achievement of the work presented here, that the width of the $X(5568)$ peak, and not only its mass, is well reproduced, fitting only one parameter in the unitarized amplitude, but, on the other hand, further discussions about the obtained results are in order. One gets a better feeling of the results if one regularizes the loop function, Eq.~\eqref{eq:GloopSubtracted}, with a sharp momentum cutoff $\Lambda$ (see Ref.~\cite{Oller:1998hw} for an explicit formula). If one performs the same fit explained before but employing this alternative regularization method, one obtains the same results (for the spectrum and the $X(5568)$ parameters) with $\Lambda = 2.80 \pm 0.04\ \text{GeV}$ for the ``cone cut'' spectrum, and $\Lambda=2.83 \pm 0.04\ \text{GeV}$ for the no ``cone cut'' spectrum. The value of this cutoff is quite large if compared with a ``natural size'' value of $\Lambda \sim 1\ \text{GeV}$. If a cutoff of this order is used, it is not possible to reproduce the spectrum nor the parameters of the claimed $X(5568)$ state, and a much broader pole at a $200\ \text{MeV}$ higher mass is produced. The fact that such a large value $\Lambda \sim 2.8\ \text{GeV}$ is necessary to reproduce the experimental information clearly points to the presence of missing channels, contributions of other sources of interactions, or to the existence of ``non-molecular'' components, such as tetraquarks. In this sense, our results would go in the directions pointed out in Refs.~\cite{Burns:2016gvy,Guo:2016nhb}, that a pure molecular state, dynamically generated by the unitarity loops, is not favoured. However, our analysis also shows that the coupling of such components to the explicit channels that we have considered is also important, apart from unavoidable ($B_s\pi$ is the decay channel), to understand the features of the peak, in particular, its width.

\begin{table}
\begin{tabular}{cccc}
Sector ($B$ or $D$)            & $J^P$ & Mass (MeV) & Width (MeV) \\ \hline\hline
$B_s\pi$, $B\bar{K}$           & $0^+$ & $5564.1 \pm 2.7$ & $27.4 \pm 1.2$ \\ 
$B^\ast_s\pi$, $B^\ast\bar{K}$ & $1^+$ & $5610.8 \pm 2.6$ & $26.8 \pm 1.3$ \\ \hline
$D_s\pi$, $DK$                 & $0^+$ & $2210.8 \pm 2.3$ & $50.1 \pm 1.4$ \\
$D^\ast_s\pi$, $D^\ast K$      & $1^+$ & $2346.7 \pm 2.3$ & $46.5 \pm 1.3$ \\ \hline 
\end{tabular}
\caption{Predictions of $X(5568)$ and partners in different spin-flavor sectors when a cutoff $\Lambda = 2.80 \pm 0.04\ \text{GeV}$ is used to regularize the loop function.\label{tab:sectors}}
\end{table}

An argument given in Ref.~\cite{Guo:2016nhb} to disfavour the molecular interpretation is that, using arguments of heavy flavor symmetry, one should expect partners of this state. In particular, in the charm sector, an $I=1$ $D_s\pi$ state (different from $D^\ast_{s0}(2317)$ \cite{Agashe:2014kda}) around $2.2\ \text{GeV}$ should be seen. Indeed, the $I=1$ matrix elements in Eqs.~\eqref{eq:pot-diag} and \eqref{eq:pot-offdiag}, computed for $B_s\pi$, $B\bar{K}$ are the same for $D_s\pi$, $DK$, with obvious masses replacements, and thus we can look for states in this other sector. By using the same cutoff $\Lambda = 2.80 \pm 0.04\ \text{GeV}$,\footnote{The use of a flavor independent cut off to respect heavy hadron flavor symmetry was invoked in Ref.~\cite{Ozpineci:2013qza}. A different method to regularize $G$ respecting this symmetry was proposed in Ref.~\cite{Altenbuchinger:2013vwa}, but in Ref.~\cite{Lu:2014ina} it was shown to be equivalent to using a common cutoff.} we find a state with mass $2210.8 \pm 2.3\ \text{MeV}$ and width $50.1 \pm 1.4\ \text{MeV}$. The prediction done here for the mass of the $D_s\pi$, $DK$ state is very similar to the one obtained in Ref.~\cite{Guo:2016nhb} using the fact that the binding energy is approximately independent of the heavy flavour. This state, in principle, has not been seen, but it is intriguing to see that a peak at $2.17\ \text{GeV}$ with a width around $50\ \text{MeV}$ is seen in the $D^+_s \pi^0$ spectrum in which the $D^{*+}_{s0}(2317)$ is seen \cite{Aubert:2006bk}. This peak is described in Ref.~\cite{Aubert:2006bk} as a reflection of the process $D^\ast_s \to D_s \gamma$, while the full distribution is additionally fitted in terms of the $D^\ast_{s0}(2317)$ decay, some background and misidentified events with unknown strengths. The possibility that this $D_s\pi$ peak be the charmed partner of the claimed $X(5568)$ state makes a reanalysis of the data in Ref.~\cite{Aubert:2006bk} advisable.

If the $X(5568)$ peak \cite{D0:2016mwd} is confirmed, the results obtained in our work can be taken farther, and predictions can also be made for the sector with heavy-flavor vector mesons, $B^\ast_s \pi$--$B^\ast\bar{K}$ and $D^\ast_s\pi$--$D^\ast K$. The vector channels have the same interaction, up to different masses, and the channels involving vector and pseudoscalar heavy mesons do not mix in this case \cite{Guo:2016nhb}, since they are in $S$-wave. Using again the same cutoff $\Lambda = 2.80 \pm 0.04\ \text{GeV}$, we predict in Table~\ref{tab:sectors} masses and widths of these $J^P = 1^+$ states. In Ref.~\cite{D0:2016mwd} the $X(5568)$ peak is interpreted as a $J^P = 0^+$ resonance, with $B_s \pi$ in $S$-wave. However, Ref.~\cite{D0:2016mwd} does not exclude the possibility that this claimed state actually decays through the chain $B^\ast_s\pi$, $B^\ast_s \to B_s \gamma$, where the low energy photon is not detected. In this case, always according to Ref.~\cite{D0:2016mwd}, it would be a $J^P = 1^+$ state and the mass of the peak would be shifted towards higher energies by an amount $M_{B^\ast_s} - M_{B_s} \simeq 50\ \text{MeV}$, while the width would remain unchanged. This would lead to $M_X \simeq 5618\ \text{MeV}$. We see in Table~\ref{tab:sectors} that such a state is also predicted with the same cutoff used in our work, and thus one would run into the same problems of interpretation we have discussed. In any case, the observation or non-observation of these resonances predicted in Table~\ref{tab:sectors} in devoted experiments would certainly bring very valuable information to unravel the present puzzle.

\paragraph{Summary.---} The D0 Collaboration has recently announced \cite{D0:2016mwd} the observation of a resonant-like peak, called $X(5568)$, in the $B^0_s\pi$ invariant mass spectrum coming from $p\!\bar{p}$ collision at $1.96\ \text{TeV}$. However, the LHCb collaboration has presented preliminary results \cite{LHCb:2016ppf} for the same spectrum, coming from $pp$ collisions, with negative results for the search of this state. We have presented the first theoretical attempt to reproduce the spectrum in which the $X(5568)$ peak has been seen. We have used an $I=1$ $B_s\pi$--$B\bar{K}$ coupled channel analysis, using an interaction potential calculated from Heavy Meson Chiral Perturbation Theory, and implementing exact unitarity. The spectrum can be well reproduced, and a pole that can be associated to the $X(5568)$ state is found, with mass and width in agreement with the one reported in the experimental analysis. However, the interpretation of this result is far from being easy, since a cutoff $\Lambda \sim 2.8\ \text{GeV}$, much larger than a ``natural value'' $\Lambda \sim 1\ \text{GeV}$, is required to reproduce the spectrum. This fact points to the presence of physical mechanisms other than the simple rescattering effects between the $B_s\pi$, $B\bar{K}$ channels, if the peak observed corresponds to a physical state.

\paragraph{Acknowledgements.---} M.~A. acknowledges financial support from the ``Juan de la Cierva'' program
(27-13-463B-731) from the Spanish MINECO. This work is supported in part by the
Spanish MINECO and European FEDER funds under the contracts FIS2014-51948-C2-1-P, FIS2014-51948-C2-2-P, FIS2014-57026-REDT and SEV-2014-0398, and by Generalitat
Valenciana under contract PROMETEOII/2014/0068.

\bibliographystyle{plain}

\end{document}